\documentstyle[12pt,epsfig]{article}

\def\aprle{\buildrel < \over {_{\sim}}}
\def\aprge{\buildrel > \over {_{\sim}}}

\begin{document}

\begin{center}

{\Large\bf COSMIC RAY MUON PHYSICS}

\vskip .6 cm

S. CECCHINI$^{1,2}$ and M. SIOLI$^{1,3}$\\

\vskip .6 cm
\footnotesize
1. INFN, Sezione di Bologna, 40126 Bologna, Italy \\
2. Istituto TESRE/CNR, 40129 Bologna, Italy \\
3. Dipartimento di Fisica dell'Universit\`a di Bologna, 40126 Bologna, Italy \\
\vskip .4 cm
{\it E-mail: cecchini@bo.infn.it, sioli@bo.infn.it}

\vskip .8 cm
{\bf Abstract}\\
\end{center}
\vskip .5 cm


We present a review of atmospheric and underground muon flux measurements.
The relevance of these data for the atmospheric neutrino flux computation
is emphasized.
Possible sources of systematic errors in the measurements are discussed,
focusing on the sea level muon data. Underground muon data are also 
reported.


\section{Introduction}
At sea level, together with neutrinos, muons are the most abundant 
particles originated by the interactions of primary cosmic rays at 
the top of the atmosphere. Due to their relative 
stability and small cross sections, these particle are able to arrive
deep underground and/or deep underwater. As a consequence, their study 
covers many aspects of cosmic ray physics. 

Recently, atmospheric muon flux measurements received attention
the context of the atmospheric neutrino anomaly
\cite{bartol,circella1}.
Because of the close relation between muon and neutrino production, it 
follows that the evaluation of the atmospheric muon flux can provide an 
important cross check on the atmospheric neutrino flux. 
Moreover, measurements of muon flux at all geomagnetic latitudes are 
crucial for the normalizazion of the calculated neutrino flux. 

Measurements performed at different altitudes (sea level, at mountain level
or in balloon born experiments) offer various advantages.
First of all, a different set of measurements at different altitude
values can provide informations about the longitudinal development of
the muon component in cosmic ray showers.
Moreover, the interpretation of data collected at the top of the 
atmosphere are not affected by the uncertainties inherent in 
particle production and propagation, since the muons (and the corresponding
neutrinos) generated in the first stages of the cascade.
Finally, the knowledge of high altitude muon data is crucial for sea level 
sub-GeV neutrinos: the corresponding sub-GeV muons originated in the
same decay processes cannot reach the sea level, considering that
the average muon energy loss in the atmosphere is of the order of 2 GeV;
we are thus forced to take data at high altitudes.

On the other hand, measurements performed at ground level offer the advantage 
of a high stability, large collecting factor and a long exposure time due 
to the relatively favourable experimental conditions. They however 
suffer of an intrinsic difficulty in interpretation, since the muons 
that arrive at sea level are the last stage of a multi-step cascade 
process. This is true, in particular, for the measurements at high 
zenith angles, near the horizon, where the intermediate and high 
energy regions of the spectrum can be analysed ($p_{\mu}$= 10-100 GeV/c).
Nevertheless, for this reason, sea level data offer the possibility to 
perform a robust check of the reliability of existing Monte Carlo codes.

Finally, underground measurements offer the possibility to extend the
energy range of the muon spectrum beyond 1 TeV. Such measurements are 
of an indirect type, but their link with the direct low-energy observations 
gives the possibility to complete the picture of muon spectra measurements 
and to cross-check the validity of the global set of data.

Most of the experiments devoted to the measurement of the muon 
momentum spectra and intensity have been carried in the '70s. 
The problem is that the results are often in disagreements with one another; 
the discrepancies are significantly larger than the experimental errors 
reported.
Recently new instruments, mainly designed for balloon experiments, 
have been developed; they are able to give detailed information 
on the muon flux at different altitudes in the atmosphere \cite{circella2}. 
Also new measurements deep underground 
\cite{enikeev,bakatanov,zatsepin,lvd,macro1} or by EAS arrays \cite{eastop} 
have added new information at very high energies.

In this paper we summarize the observations of the muon flux at 
sea level and deep underground and discuss some of the systematics
connected with such measurements.
For more complete discussion one can refer to the recent papers 
\cite{knapp,bugaev} and to books \cite{gaisser,hayakawa}.


\section{Atmospheric muon production and propagation}
Secondary muons are mainly produced in the decays of secondary 
mesons, mostly $\pi^{\pm}$ and $K^{\pm}$. 
The most important decay channels, and their respective decay 
probabilities, are:
\[a) \hskip 20pt \pi^{\pm}\rightarrow \mu^{\pm}\nu_{\mu} \hskip 40pt\sim 100\%\]
\[b) \hskip 20pt K^{\pm} \rightarrow \mu^{\pm} \nu_{\mu}\hskip 40pt\sim 63.5\%\]
in which the produced muons take on the average 79\% and 52\% of the energy 
of the $\pi^{\pm}$ and $K^{\pm}$, respectively. 

The contribution of $K$ decays to muon production is a function of 
the energy and ranges from $\sim$ 5\% at low energies to an asymptotic 
value of $\sim$ 27\% for E $\aprge 1$ TeV. At very high energies a small 
contribution arises from charmed particles. 
The analytical form of the muon production spectrum at a given height in
the atmosphere can be derived by folding the two-body decay kinematics of
the parent mesons with their production spectrum. The latter is
generally expressed in terms of the so called ``spectrum weighted''
moments
\begin{equation}
Z_{p\pi^{\pm}} = \int_{0}^{1} {x^{\gamma} \frac{dN_{p\pi^{\pm}}}{dx} dx}
\label{swm}
\end{equation}
where $dN_{p\pi^{\pm}}/dx$ is the pion production spectrum 
($x=E_{\pi}/E_{p}$ and $\gamma$ is the differential primary 
spectral index). 
A similar expression can be obtained for kaons.
In general, the development of the meson and muon components
in the atmosphere depends on the energy range we are considering.
The competition between interaction and decay of the particles
plays a crucial role and the relative importance of the two
processes depends on the energy. We can distinguish three different 
energy regions in the muon spectrum:\\

\vskip 0.2 cm

\par a) $E_{\mu} \gg \epsilon_{\pi,K}$ , where $\epsilon_{\pi}$ = 115 GeV 
and $\epsilon_{K}$ = 850 GeV are the critical energy beyond which meson 
reinteractions cannot be neglected.
This is the typical muon energy range studied by underground detectors
or by ground based experiments looking at high inclined directions.
In this case, the meson production spectrum have the same power law
dependence of the primary cosmic rays, but the rate of their decay
has an extra $E^{-1}$ dependence with respect to the primary and meson
spectrum (a consequence of the Lorentz time dilatation).
The muon (and hence neutrino) flux takes the form:
$dN/dE_{\mu}=E_{\mu}^{-(\gamma+1)}$, and a zenith dependence
$dN/dcos\theta \propto (cos\theta)^{-1}$.
It should be noted that, in this energy region, the enhancement of the 
$K^{\pm}$ contribution to the secondary lepton production is particularly 
important in the neutrino flux calculation, as a consequence of the two-body
decay kinematics \cite{gaisser_calgary}. 
This last remark does not hold for muons, for which the limited knowledge
of meson production in this energy range is not so crucial as for
neutrinos.

\vskip 0.2 cm

\par b) $\epsilon_{\mu} \aprle E_{\mu} \aprle \epsilon_{\pi,K}$ , where 
$\epsilon_{\mu} \simeq$ 1 GeV. In this energy range, almost all
the mesons decay, and the muon flux has a power law dependence with
the same spectral index of the parent mesons (and hence of the primary
cosmic ray, in the assumption of complete Feynman scaling validity)
and is almost independent on the zenith angle.
A compact form which expresses the low and high energy regions is 
\cite{gaisser}:
\begin{equation}
\frac{dN_{\mu}}{dE_{\mu}}(E_{\mu},\theta) \simeq 0.14\,E_{\mu}^{-2.7}
\left[\frac{1}{1+\frac {1.1 E_{\mu}cos\theta}{\epsilon_{\pi}}}+
\frac{0.054}{1+\frac {1.1 E_{\mu}cos\theta}{\epsilon_{K}}}\right]
\label{eq:muflux}
\end{equation}

\vskip 0.2 cm

\par c) $E_{\mu} \aprle \epsilon_{\mu}$ . In this case, muon decays 
and the energy losses in the atmosphere cannot be neglected. Moreover, 
geomagnetic latitude and solar modulation now play an important role 
being the primary cosmic ray energy $E_{p} < 20$ GeV.

We stress again the relevance of muon flux measurements for the 
knowledge of the neutrino flux. In principle, sea level neutrino
flux computation can be derived directly from muon flux measurements 
high in the atmosphere ($X <$ 37 $g/cm^{2}$) \cite{perkins}.
This approach gives good results, but only a complete Monte Carlo simulation
can take into account second order effects.
The main ingredients in Monte Carlo calculations (and the main
sources of systematics) of atmospheric lepton production are the input 
primary cosmic ray spectrum and a detailed description of secondary 
multiparticle production in the atmosphere.
The primary cosmic ray composition plays an important role only in
the very high energy range; the composition is dominated by protons
and $\alpha$ particles at energies below 100 GeV.

Among various Monte Carlo codes now available, we recall \cite{bartol}
and \cite{honda} which are the ones used to interpret the atmospheric
neutrino anomaly, and the new code based on the FLUKA interaction model 
\cite{battista} which takes into account 3-dimentional effects of 
secondary propagation in the atmosphere.

The comparison between the Monte Carlo evaluation of the muon
flux at different altitudes and at sea level with the existing
muon flux measurements constitutes one of the most powerful benchmark
to assess the validity of the simulations.


\section{Atmospheric muon flux measurements}
Measurements of the absolute intensity, energy spectrum and 
positive-to-negative ratio of muons have been carried out many times 
in the past. 
Most of these observations were made at sea level and 
few at different mountain altitudes with counter telescopes separated 
by absorbers (Pb, Fe) and magnetic spectrometers. More recently, 
with the development of superconducting magnet, it has been possible 
to operate spectrometers also on board of balloons
\cite{heat,caprice,imax,bess,mass} 
which led to accurate measurements at different levels in the atmosphere.

Here we will consider mainly ground-based measurements and those made 
with detectors on balloons near the ground level or very close to it.
The relevant quantities that can be directly measured and will be
discussed here, are:
\par - absolute muon intensity
\par - muon momentum spectrum
\par - charge ratio

\subsection{Absolute intensity measurements}
The vertical muon intensity at sea level is a quantity which varies 
with the geomagnetic latitude, altitude, solar activity and 
atmospheric conditions.

The geomagnetic field tends to prevent low energy cosmic rays from 
penetrating through the magnetosphere down to the Earth's atmosphere. 
At any point on the Earth one can define a threshold or cut-off 
rigidity, Pc, for cosmic rays arriving at a particular zenith and 
azimuth angle \cite{rossi2}. 

\begin{figure}[thb]
\begin{center}
\epsfig{file=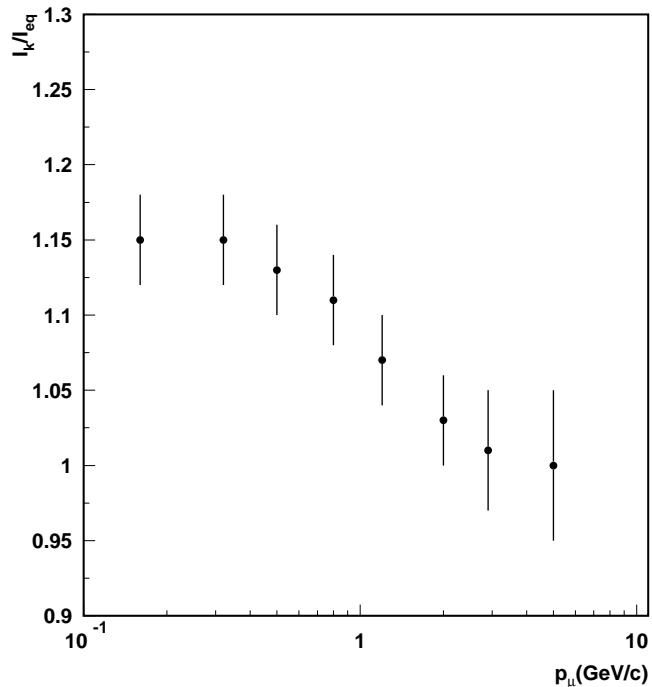,width=9cm}
\end{center}
\vskip -1.0 cm
\caption{\em The latitude effect on the integral muon intensity at sea
level. It is shown the ratio between the intensity measured at Kiel 
$I_{K}$ (Pc = 2.3 GV), and the intensity measured near the equator 
$I_{eq}$ (Pc = 14.1 GV), with the same instrument $^{26}$.
\label{f:fig1}}
\end{figure}

Primary nuclei having lower rigidity are excluded by 
the action of the geomagnetic field and do not contribute to 
production of secondaries in the atmosphere. The cut-off values range 
from less than 1 GV near the geomagnetic poles to about 16 GV for 
vertical particles near the equator \cite{smart}.
It results that geomagnetic effects are important for sea level
muons up to about $\sim$ 5 GeV (Fig. \ref{f:fig1}).
The effect is larger at higher altitudes; Conversi \cite{conversi}
found that the vertical flux of muons with momentum around 0.33 GeV/c
at latitude 60 deg was 1.8 times higher with respect to the flux 
at the equator.

Moreover, as cosmic ray primaries are predominantly positively charged
particles, the flux and spectra in the East and West directions differ 
up to energies of about 100 GeV; the intensity from the West is stronger 
than that from the East. This effect increases with altitude.

\begin{figure}[thb]
\begin{center}
\epsfig{file=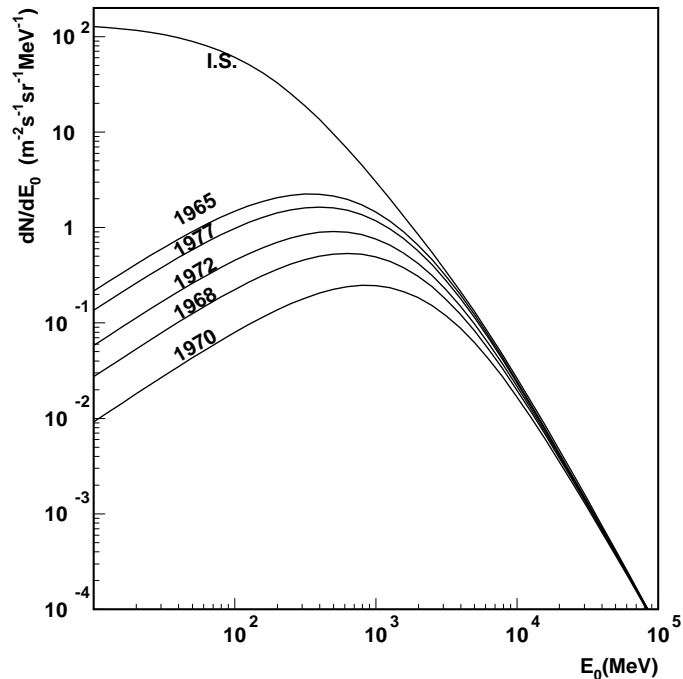,width=9cm}
\end{center}
\vskip -1.0 cm
\caption{\em Approximate solutions (the so called 
force-field solution) $^{27,28}$ that fit the observations of 
the primary proton spectrum at the top of the atmosphere at different years. 
Also shown is the assumed Interstellar Spectrum. Notice the extension of the 
solar modulation effect.
\label{f:fig2}}
\end{figure}

In addition, the primary cosmic ray spectrum at the top of the atmosphere
changes with the 11 year solar cycle as the configuration of the 
Interplanetary Magnetic Field (IMF) varies. It results that the cosmic ray
flux is significantly ``modulated'' up to energies of about 20 GeV 
(Fig. \ref{f:fig2}).

\begin{figure}[thb]
\begin{center}
\epsfig{file=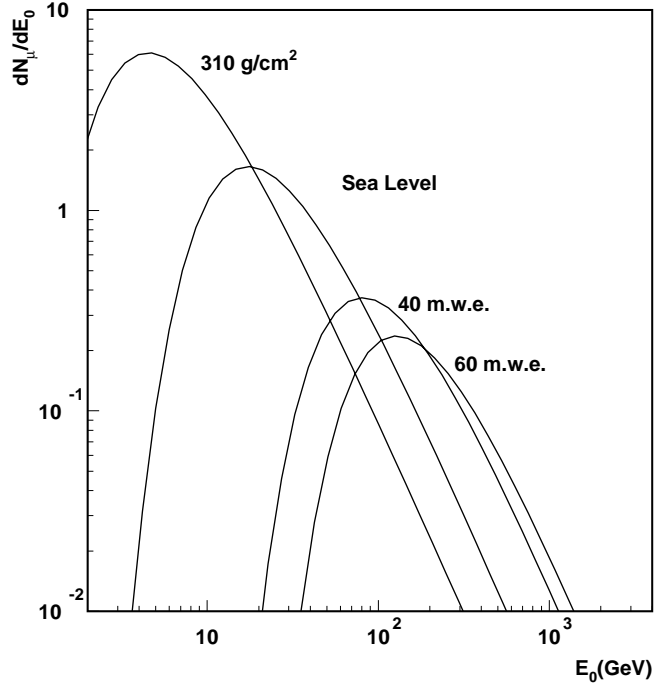,width=9cm}
\end{center}
\vskip -1.0 cm
\caption{\em Differential response functions for muons detected at
different depths, after Mathews $^{29}$
\label{f:fig3}}
\end{figure}

In order to estimate how these changes in the primary spectrum influence
the counting rate of a muon detector it is necessary to know the
"differential response curve" \cite{gaisser}. Its shape varies 
significantly with the depth of observations, see Fig. \ref{f:fig3}.
Their detailed calculations depend on the properties of nuclear cascades 
in the atmosphere; more precise descriptions can be found in
\cite{gaisserj,murakami}
At the standard momentum of 1 GeV/c and at high latitudes the modulation 
is 7\% and 4.5\% for the differential and the integral fluxes, 
respectively \cite{allkofer2}.

\begin{figure}[thb]
\begin{center}
\epsfig{file=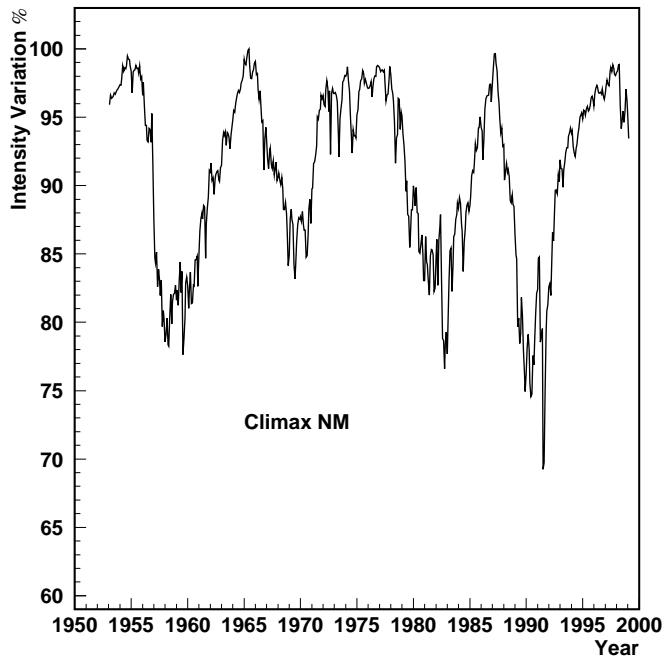,width=9cm}
\end{center}
\vskip -1.0 cm
\caption{\em Observed time variation in the monthly counting rate of the 
Climax Neutron Monitor (Pc = 3.03 GV) normalized to 1965. 
The 11 year variation associated with solar sunspot cycle is easily 
recognizable. 
Pronounced minima occur during sunspot maxima. 
A complete record of the muon intensity variations during the same 
period is not available.
\label{f:fig4}}
\end{figure}

So in making a comparison of muon observations at low energies (less than
20 GeV) it is very important to 
know the year and the location the measurements were made. 
Figure \ref{f:fig4} shows the neutron monitor counting rate recorded by a 
middle latitude station since 1953. No continuous recording of the same kind 
exists for muon monitors. By the comparison of the peak to peak variations
during the interval 1965-1972 one can estimate that the total muon flux
changes are usually a facto 3 to 5 smaller than the observed neutron flux
variations \cite{allkofer2,dorman1}

Finally, changes in pressure and, particularly, temperature above the 
instrument up to the point of muon production by pions and kaons,
produce variations of different amplitude in different energy range. 
The most conspicuous for muons at higher energies is the seasonal 
variation \cite{dorman2,barrett,macro2} for which the results reported in
the following have not been corrected for. 
 
Classical definition of the hard component \cite{rossi1} is related 
to penetration characteristics, it is to say the capability of 
crossing 167 $g/cm^{2}$, equivalent to roughly 15 cm of Pb. 
As a matter of fact this component is made of of muon with momenta 
$p_{\mu} > 0.32$ GeV/c and less than 1\% are protons, neutrons, 
electrons and pions. 

Let us distinguish between vertical and horizontal 
integral muon fluxes. These latter are made in order to extend the 
range of the former beyond several tens of GeV/c, but usually they do 
not give absolute values of the vertical muon intensity. For this
reason we will not discuss them here.

\begin{figure}[thb]
\begin{center}
\epsfig{file=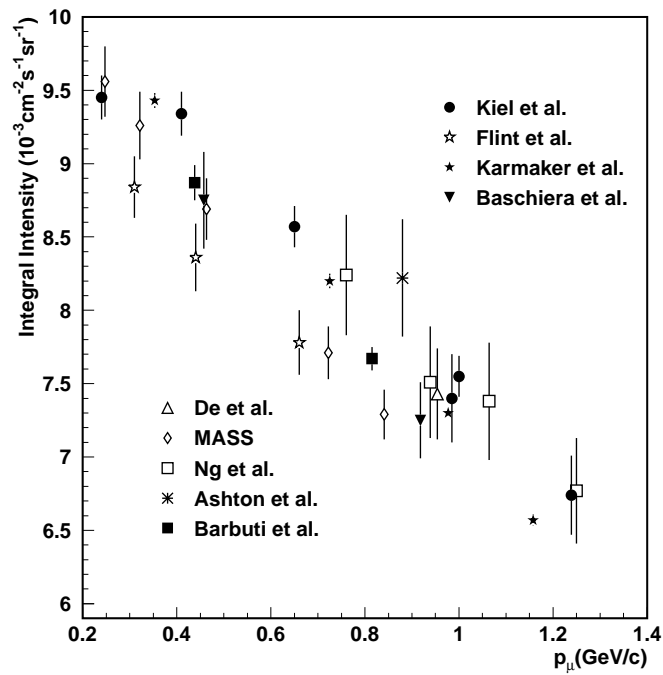,width=9cm}
\end{center}
\vskip -1.0 cm
\caption{\em Integral momentum spectrum of low energy muons at sea level.
Data are taken from 
Kiel $^{26,32}$, 
Flint et al. $^{39}$, 
Karmakar et al. $^{45}$, 
Baschiera et al. $^{41}$, 
De et al. $^{46}$,
MASS $^{44}$,
Ng et al. $^{40}$,
Ashton et al. $^{42}$, 
Barbuti et al. $^{43}$.
\label{f:fig5}}
\end{figure}

The first measurement of the integral vertical intensity was made by 
Greisen \cite{greisen} at latitude $50^{o}$ and altitude 259 m a.s.l. 
(corresponding to 1007 $g/cm^{2}$) who found the value: 
$0.83 \times10^{-2} \pm  1\%$ $cm^{-2} s^{-1} sr^{-1}$. 
Rossi \cite{rossi1} noticed that this value needed to be 
corrected in order to account for showering and scattering 
of particles inside the apparatus.
Successive measurements led to higher values.
The observations made at different latitudes and during different years 
are presented in Fig. \ref{f:fig5}.
Most measurements were made at high latitudes 
\cite{flint,ng,baschiera,ashton,barbouti,depascale},
and only the few at low latitudes \cite{karmakar,de}; 
were corrected for the geomagnetic effect.
No corrections have been made for the solar modulation
effects; the measurements are essentially grouped in the period 1967-1977,
with one in 1998 \cite{depascale}. 
The agreement between the measurements is fairly good 
(all the data within 10\%) and one has to take into account that the
largest contribution to the deviations are the 
systematic errors due to incorrect knowledge of the acceptance, 
efficiency of the counters and correction for the multiple scattering.

For energies $E_{\mu} > 1$ TeV direct measurments of the muon flux 
were made at highly inclined directions using large magnetic 
spectrometers \cite{muraki,matsuno,jokisch}, large emulsion chambers 
\cite{ivanova} 
and EAS arrays \cite{eastop}. For the discussion of the results 
we refer to the recent paper \cite{bugaev}.

\subsection{Momentum spectra}
These spectra have been measured many times for moments up to 
$\sim$ 100 TeV/c. 
Magnetic spectrometers are mainly used at low and intermediate energies, 
while observations at high energies are made close to the horizontal 
directions at ground level or deep underground. 
The latter are indirect measurements, since the ground level spectra 
have to be extracted from underground data. 
We will consider here only ground level and underground observations.

\subsubsection{Ground level measurements}
Direct measurements of momentum spectra for $p_{\mu} < 1$ TeV/c are
important for the comparison of nuclear cascade models with 
available data. Furthermore by extending the model results to higher
energies one can hope to be able to evaluate prompt muon production
and/or charm production.
In the momentum region 10 GeV/c - 1 TeV/c:
a) the production spectrum of the charged pions cannot be represented 
by a power law but has a maximum at an energy that depends on both 
the altitude and the latitude
b) the energy loss and the decay of muons must be properly considered.
In order to join low to high momentum spectra it is important to have
single experiments that cover the widest energy range.

\begin{figure}[thb]
\begin{center}
\epsfig{file=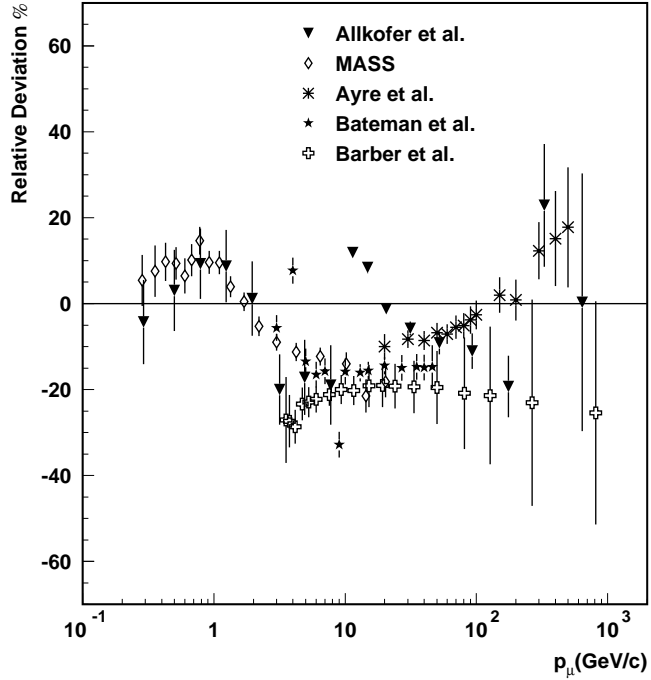,width=9cm}
\end{center}
\vskip -1.0 cm
\caption{\em Relative deviation of the differential muon spectra with
respect to the Kiel fit $^{51}$. The data are taken from 
Allkofer et al. $^{51}$, 
MASS $^{44}$, 
Ayre et al. $^{52}$, 
Bateman et al. $^{53}$,
Barber et al. $^{54}$.
\label{f:fig6}}
\end{figure}

To better see the differences between the sea-level spectra
we plot in Fig. \ref{f:fig6} the percentage deviations of 
the data from the best fit spectrum obtained by  \cite{allkofer1}.
Notice that even if individual errors are small (however increasing
with momentum due to decreasing number of detectable particles and
to the maximum detectable momentum), deviations up to $\pm$ 20\% 
are observed probably because of systematic effects.

\subsubsection{Underground measurements}

From underground muon intensity measurements, informations about 
sea-level muon spectra can be obtained using different procedures 
(see for example \cite{bakatanov,zatsepin}).
Here and in the following, we assume a standard procedure applied
form large area underground experiment \cite{macro1,lvd}.
The vertical muon intensity, for a given direction $\theta,\phi$ and 
a corresponding rock slant depth $h$ can be expressed as:

\begin{equation}
I^{V}_{\mu}(h,\theta,\phi)=\left(\frac{1}{\Delta T}\right)
\frac{\sum_{i}N_{i}m_{i}}{\sum_{j}\Delta\Omega_{j}A_{j}\epsilon_{j}
/cos\theta_{j}}
\label{eq:i_h}
\end{equation}

where $\Delta T$ is the total livetime of the experiment, 
$N_{i}$ is the number of detected events with multiplicity 
$m_{i}$ in the angular bin $\Delta\Omega_{j}$, $A_{j}$ and 
$\epsilon_{j}$ are, respectively, the geometrical and intrinsic 
acceptance of the detector.
The relation between the measured $I^{V}_{\mu}(h)$ and the sea-level
muon spectrum can be expressed as:

\begin{equation}
I^{V}_{\mu}(h)=\int_{0}^{\infty}\frac{dN_{\mu}}{dE_{\mu}d\Omega}
P(E_{\mu},h)dE_{\mu} ,
\label{eq:unfolding}
\end{equation}

where $P(E,h)$ is the muon survival probability function determined
via Monte Carlo. Assuming for the sea-level muon spectrum an expression
of the form (\ref{eq:muflux}), leaving as free parameters the muon
spectral index and a normalization constant, it is possible 
it is possible to unfold sea level muon spectrum from the measured 
absolute muon intensity.

\begin{figure}[thb]
\begin{center}
\epsfig{file=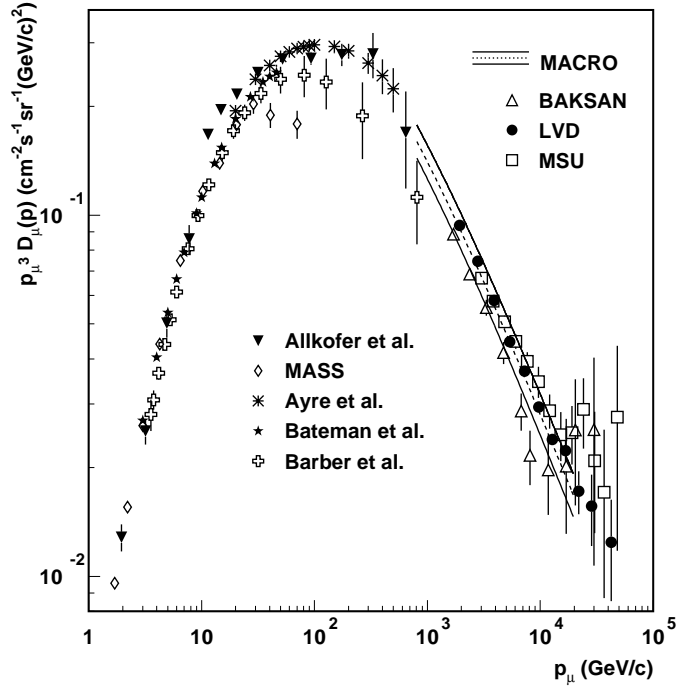,width=9cm}
\end{center}
\vskip -1.0 cm
\caption{\em Vertical differential momentum of muons at sea level.
Direct data extend up to 1 TeV and are taken from 
[allkofer, depascale, barber, bateman, ayre]. Indirect data obtained from
underground observations for E $> $1 TeV/c are taken from 
Baksan $^{5}$,
LVD $^{7}$,
MSU $^{6}$,
and MACRO fit $^{8}$, with the two parallel lines indicating the range
allowed by the errors of the fit parameters.
\label{f:fig7}}
\end{figure}

In Fig. \ref{f:fig7} are reported the results of the fit of 
MACRO data \cite{macro1} together with LVD \cite{lvd}, 
MSU \cite{zatsepin} and Baksan \cite{bakatanov} data.
Data are presented multiplied by factor $p^{3}$ to better observe
the variation of the spectrum in the whole energy region and
to strengthen a possible flattening in the tail of the spectrum 
due to charm production. 
The statistics is still too poor to allow any definite assessment
on the existence of this effect at energies $>$ 10 TeV/c.

In indirect measurements, accurate estimates of the systematic
errors are needed. The main sources of systematics in (\ref{eq:unfolding}) 
are the knowledge of the rock density overburden and the 
treatment of hard processes in the energy loss of muons in the rock. 
In the MACRO fit, for example, their overall contribution has been estimated 
to be $\sim$ 5\% and 3\% in the determination of the normalization constant 
and muon spectral index respectively.

\subsection{Charge ratio}

\begin{figure}[thb]
\begin{center}
\epsfig{file=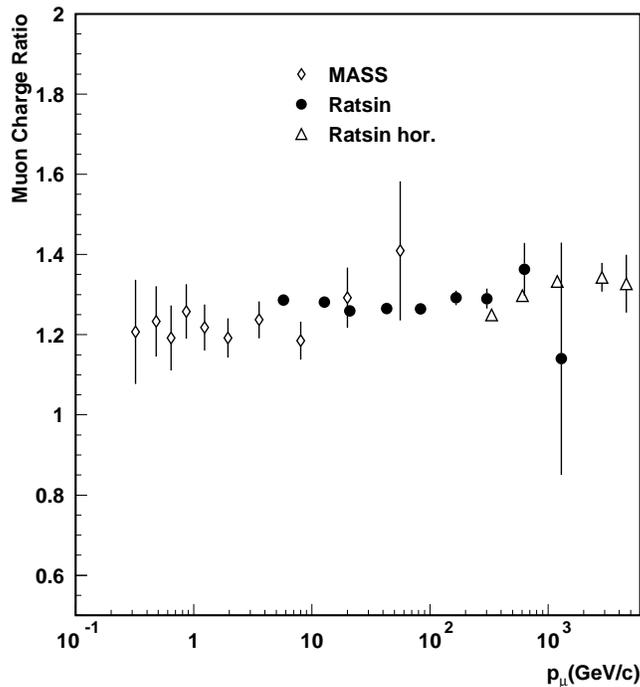,width=9cm}
\end{center}
\vskip -1.0 cm
\caption{\em Ratio $\mu^{+}/\mu^{-}$ at sea level. The data have been
taken from MASS~$^{44}$, Rastin $^{55}$.
\label{f:fig8}}
\end{figure}

In the primary cosmic rays there is an excess of positively charged
particles (protons) with respect to the total number of nucleons.
This excess is transmitted via nuclear interactions to pions and further to
muons. By assuming that the primary composition is constant in the energy
range considered, this ratio will remain constant with the exception 
of high energies, where the contribution from kaons starts to become sizeable.
The muon charge ratio is expected to increase also with zenith angle as the
depth is increasing and likewise the energy of the primaries that produce
muons of a given momentum at ground.
This quantity is important to study nucleon-nucleon interactions, 
composition and kaon contribution. 
Magnetic spectrographs are used for determining this ratio.
Because of systematic effects in the momentum measurement the values are 
usually much spread out. 
Moreover limited statistics at high energy makes it difficult to
appreciate the energy dependence.
We report in Fig. \ref{f:fig8} only the recent data from Mass 
\cite{depascale} at lower momenta, and the two compilations made by 
\cite{rastin}.
It is clear more measurements with longer exposures are still needed.

\section{Conclusions}
Since atmospheric muons and neutrinos are generated in 
the same processes, the accuracy of the neutrino flux 
calculation can be improved by forcing the poorly known input 
parameters of the cascade model to fit the data on the muon flux.

However, the data are still not sufficient for this purpose, since 
several sea level measurements of the vertical muon flux are in poor 
agreement with one another, even though each experiment has 
typically very good statistics. 

Disagreement between the results of different experiments 
are present even if the quoted errors are relatively small 
in the majority of the experiments. It indicates the existence of 
significant systematic errors in some experiments by as much as 30-35\% 
at momenta from 10 to 1000 GeV/c.



\end{document}